\documentclass[aps,pre,twocolumn,nofootinbib]{revtex4-1}

\usepackage{graphicx}
\usepackage{latexsym}
\usepackage{amsmath}
\usepackage{amsfonts}
\usepackage{amssymb}
\usepackage{color}

\usepackage{units}

\newcommand{\be}{\begin{equation}}
\newcommand{\ee}{\end{equation}}
\newcommand{\bea}{\begin{eqnarray}}
\newcommand{\eea}{\end{eqnarray}}

\begin{document}
\title{Non-Equilibrium Interaction Between Catalytic Colloids:  Boundary Conditions and Penetration Depth}

\author{Alexander Y. Grosberg}
\affiliation{Department of Physics and Center for Soft Matter Research, New York University, 726 Broadway, New York, NY 10003 USA}
\author{Yitzhak Rabin}
\affiliation{Department of Physics and Institute of Nanotechnology and Advanced Materials,
Bar-Ilan University, Ramat Gan 52900, Israel}

\date{\today}

\begin{abstract}
 Spherical colloids that catalyze the interconversion reaction $A \leftrightharpoons B$ between solute molecules $A$ and $B$ whose concentration at infinity is maintained away from equilibrium effectively interact due to the non-uniform fields of solute concentrations. We show that this long range $1/r$ interaction is suppressed via a mechanism that is superficially reminiscent but qualitatively very different from electrostatic screening: catalytic activity drives the concentrations of solute molecules towards their equilibrium values and reduces the chemical imbalance that drives the interaction between the colloids. The imposed non-equilibrium boundary conditions give rise to a variety of geometry-dependent scenarios; while long-range interactions are suppressed (except for a finite penetration depth) in the bulk of the colloid solution in 3D, they can persist in quasi-2D geometry in which the colloids but not the solutes are confined to a surface, resulting in the formation of clusters or Wigner crystals, depending on the sign of the interaction between colloids.
\end{abstract}

\maketitle

\section{Introduction}

To build a macroscopic machine capable of directly utilizing chemical energy to perform mechanical work, bypassing heat, is a long standing and unresolved engineering challenge.  At the same time, on the macromolecular or colloidal scale, this is routinely done by molecular motors moving on a solid substrate  \cite{Motors_Review_RevModPhys.69.1269} or by colloidal swimmers moving through a fluid \cite{ColloidalSwimmers_EBBENS201614}.  In the latter case, mechanical motion is usually achieved by diffusiophoresis, i.e., the drift of a colloidal particle (or a liquid droplet) in a solvent,  induced by gradients in the concentration of chemical species (solute) \cite{Anderson1989, Brady2011, Paustian2015, Sear2017}. The phenomenon is driven by short-range interactions between the surface of the particle and the solute molecules which result in different energies of a solute molecule close to the surface of the particle and away from it and, depending on the sign of the interaction, it leads to the motion of the particle along or opposite to the direction of the concentration gradient. Recently, diffusiophoresis has been proposed as a non-equilibrium, non-motor protein  mechanism for metabolism-dependent transport of protein filaments, plasmids, storage granules, and foreign particles of different sizes in cells \cite {Parry2014, Sear2019}. Related cross-diffusion and chemotaxis effects \cite{Vanag2009}   have been also implicated in the aggregation of enzymes and the formation of metabolons in regions of high substrate concentrations \cite{Zhao2018}.  Under the name ``chemically (or phoretically) active matter'' these systems attracted much attention from theorists in recent years.  A far reaching phenomenological theory was developed by Ramin Golestanian with co-authors \cite{PhysRevLett.108.038303, PhysRevLett.112.068301, PhysRevE.89.062316, PhysRevE.91.052304, PhysRevLett.123.018101, Saha_2019, Golestanian_Review_2019, Nasouri_Golestanian_2020} and in a number of other works \cite{PhysRevLett.115.258301, PhysRevE.81.046311, Oshanin_2017} (reviewed in \cite{Golestanian_Review_2019}).

One simple way to create solute concentration gradients is to have colloidal particles catalyzing the reaction $A \leftrightharpoons B$ between substrate $A$ and product $B$ molecules, provided that substrates are supplied to the system, while products are washed away.  An interesting observation about such a system is that concentration gradients typically decay as $1/r$ with distance, thus leading to effective interactions which are long-ranged and reminiscent of electrostatics or gravity \cite{PhysRevE.89.062316, PhysRevE.81.046311, Oshanin_2017}.  This realization leads to prediction of a plethora of beautiful and unusual states of this ``phoretically active matter'' \cite{PhysRevE.89.062316}.

We here want to revisit that same system in order to clarify one aspect of it, which is the following.  {\it Whenever there is a catalyst that accelerates chemical transformation of $A$ (``fuel'') to $B$ (``exhaust'') molecules, $A \rightharpoonup B$, it accelerates also the reverse reaction, $B \rightharpoondown A$; in other words, it accelerates relaxation to equilibrium}.  This fact has interesting consequences for the analog of Debye-H\"{u}ckel electrostatic screening in systems of catalytic colloids.  Specifically, in electrostatics, the field that is being screened is, of course, the electric field, or potential.  What is screened in our chemical system?  We shall show that it is the field of chemical imbalance that measures the deviation from chemical equilibrium, $\psi(\mathbf{r}) \equiv k_{\rightarrow}  c_{A}(\mathbf{r}) - k_{\leftarrow} c_{B}(\mathbf{r})$, where $c_{A}(\mathbf{r})$ and $c_{B}(\mathbf{r})$ are local concentrations of solute components, while $k_{\rightarrow}$ and $k_{\leftarrow}$ are corresponding catalytic rate constants.  For instance, in a canonical example, when there is a large crowd of catalytic particles confined in an osmotic bag permeable for fuel $A$ and exhaust $B$ molecules, but not permeable for catalytic particles, and even if chemical imbalance is maintained outside by supplying $A$ and removing $B$, the chemical imbalance field $\psi(\mathbf{r})$ penetrates into the crowd only by a finite distance and decays exponentially beyond that distance.  Deep inside the crowd of catalysts both $A$ and $B$ are present, but in chemical equilibrium.  This main point of our work has some important consequences which we will discuss at the end.

The plan of this article is as follows.  To make the work self-contained and to establish the notations, we rederive some of the well-known results \cite{PhysRevE.89.062316} about concentration profiles around a single catalyst  and about interactions between two catalysts in section  \ref{sec:combined}.  This section contains no new results and serves mostly pedagogical purposes, except that unlike previous authors, we never omit the fundamentally important reverse catalytic reaction.  The crowd of catalysts, screening \cite{Debye_Screening_1923}, Wigner crystals \cite{Wigner_PhysRev.46.1002}, and clusters of catalytic colloids, are considered in section \ref{sec:many_catalysts}.

\section{The $1/r$ interaction between catalytic colloids}\label{sec:combined}


Let $c_{A}^{\infty}$ and $c_{B}^{\infty}$ be the respective concentrations (molecules per unit volume) far away from the catalysts.  We will assume that the energy barrier for the interconversion reaction $A \leftrightharpoons B$ is sufficiently high so that, {\it in the absence of catalysts}, the system can be maintained indefinitely out of equilibrium and therefore $c_{A}^{\infty}$ and $c_{B}^{\infty}$ are externally controlled parameters.

Consider first a single spherical particle of radius $R$ which can catalyze the reaction $A \leftrightharpoons B$ on its surface by reducing the energy barrier to a value comparable to $k_{B}T$. Assuming for simplicity that concentrations are sufficiently small, the steady state rate (current) of catalytic reaction can be written as
\begin{equation}  J = v   k_{\rightarrow} c_{A}(R)  - v k_{\leftarrow}c_{B}(R)  \ , \label{eq:current_general} \end{equation}
with $k_{\rightarrow}$ and $k_{\leftarrow}$ forward and backward rate constants, $ c_{A}(R)$ and $c_{B}(R)$ the concentrations of $A$ and $B$ species at the surface of the catalyst, and $v$ the volume where reaction takes place (for instance, if catalysis occurs uniformly along the spherical surface, then $v = 4 \pi R^2 d$, with $d$ a molecular length scale).  As we stated, eqn (\ref{eq:current_general}) is valid only for sufficiently small concentrations of $A$ and $B$, otherwise the catalyst gets ``clogged'' and a non-linear Michaelis-Menten reaction rate has to be used, as it was done in \cite{PhysRevE.89.062316}. However, for our purpose, it is important to have both forward and backward reaction taken into consideration, that at large concentrations would require using the so-called reversible Michaelis-Menten kinetics \cite{Reversible_MichaelisMenten_1, Reversible_MichaelisMenten_2} which was not done in  \cite{PhysRevE.89.062316}.  Because of the dramatic simplification, we stay with the linear relation (\ref{eq:current_general}). Since solute particles $A$ and $B$ have to be delivered to and from the catalyst surface by diffusion, their steady state concentration profiles must be found from the appropriate diffusion equation.  For a spherically symmetric catalyst, the concentration fields of solutes $A$ and $B$ are spherically symmetric as well:
\begin{equation}  c_{A}(r)  =  c_{A}^{\infty} - \frac{J}{4 \pi D_{A} r} \ ;  \ \  c_{B}(r)  =  c_{B}^{\infty} + \frac{J}{4 \pi D_{B} r} \ ,  \label{eq:concentrations_8} \end{equation}
where $D_{A}$ and $D_{B}$ are the corresponding diffusion coefficients.  Plugging these expressions (at $r=R$) back to eqn (\ref{eq:current_general}) which serves as a boundary condition for the diffusion equation, produces an equation for the current $J$ with the solution
\begin{equation} \frac{J}{v} = \frac{k_{\rightarrow} c_{A}^{\infty} - k_{\leftarrow} c_{B}^{\infty}}{1+ \frac{v}{4\pi R} \left[ \frac{k_{\rightarrow}}{D_{A}} +  \frac{k_{\leftarrow}}{D_{B}}  \right]} \ . \label{eq:current_5} \end{equation}
This result is easily generalized for the case when several species of $A_i$ and $B_j$ are present.

Current $J$ (\ref {eq:current_5}) vanishes in thermal equilibrium, since the equilibrium concentrations $c_{A}^{\mathrm{eq}}$ and $c_{B}^{\mathrm{eq}}$ obey the detailed balance condition, $k_{\rightarrow} c_{A}^{\mathrm{eq}}= k_{\leftarrow} c_{B}^{\mathrm{eq}}$.  In this sense, the quantity in the numerator of formula (\ref{eq:current_5}) characterizes the degree of chemical imbalance which drives the process, and which can be governed by energy (if energy of a fuel molecule is larger than that of exhaust), or by entropy (if $c_{A}^{\infty} > c_{B}^{\infty}$), or by any combination of the two.

We now consider two catalytic spheres, some distance $r$ apart, such that $r \gg R$; the catalyst spherical symmetry assumption will be relaxed later on.  Because of the short-range interactions between solute molecules $A$ and $B$ and the catalyst, and because steady state concentrations of $A$ and $B$ are non-uniform in space, the energies of these two spheres depend on the distance $r$ between them, i.e., there is an interaction force between them.  This problem can be treated, in the first approximation (see Supplementary Material \cite{SupMat1}), by imagining one particle located in the origin, while the other particle, positioned at distance $r$ away, interacts with unperturbed concentration fields $c_{A}(r), \ c_{B}(r)$ eqn (\ref{eq:concentrations_8}) created by the first. 
Expanding the surface energy of a sphere in small concentrations $c_{A}$ and $c_{B}$ at the sphere surface, as $\sigma \simeq \sigma_0 + c_{A}(r) \sigma^{\prime}_{A} + c_{B}(r) \sigma^{\prime}_{B}$ (where prime signs indicate partial derivatives of surface tension with respect to the corresponding concentration), we write distance-dependent part of energy for two spheres as follows:
\begin{equation} \frac{E}{4 \pi R^2} = \sigma^{\prime}_{A} \left[ c_{A}(r) - c_{A}^{\infty} \right] + \sigma^{\prime}_{B} \left[ c_{B}(r) - c_{B}^{\infty} \right] \  . \end{equation}
For brevity, we again drop the generalization for the case of several species $A_i$ and $B_j$.  The constant ($r$-independent) $c_{A}^{\infty}$ and $c_{B}^{\infty}$ terms are subtracted such that this energy vanishes when two droplets are infinitely far.  Plugging in the concentration profiles eqn (\ref{eq:concentrations_8}), the force on each sphere is
\begin{equation} \frac{f}{4\pi R^2} = \frac{J}{4 \pi r^2} \left[ \frac{\sigma^{\prime}_{B} }{D_{B}} - \frac{ \sigma^{\prime}_{A} }{D_{A}}\right] \ . \label{eq:Force_droplets_compact_1} \end{equation}
where the current $J$ is given by eqn (\ref{eq:current_5}) (we have neglected the hydrodynamic interaction contribution to the force, $f = - \nabla E$; see \cite{Nasouri_Golestanian_2020}). This force depends on the distance as $1/r^2$ i.e., it is a long-range interaction similar to gravitational and Coulomb forces, as it was pointed out in \cite{PhysRevLett.108.038303, PhysRevLett.112.068301, PhysRevE.89.062316, Saha_2019, PhysRevE.81.046311, Oshanin_2017}.  Furthermore, the force is proportional to $J$ -- the chemical rate (or current) of interconversion of $A$ to $B$, which emphasizes that the entire phenomenon is of non-equilibrium nature.  It is driven by the supply of fuel $A$ molecules as well as removal of exhaust $B$ molecules at infinity.

A word of caution is in order about our usage of equilibrium surface tension $\sigma$ and its derivatives $\sigma^{\prime}_{A}$ and $\sigma^{\prime}_{B}$ in this decidedly non-equilibrium context.  In fact, it is well justified by the assumption that colloidal catalytic particles are much larger and move much slower than the solute molecules $A$ and $B$.

\begin{figure}
  \centering
  \includegraphics[width=0.3\textwidth]{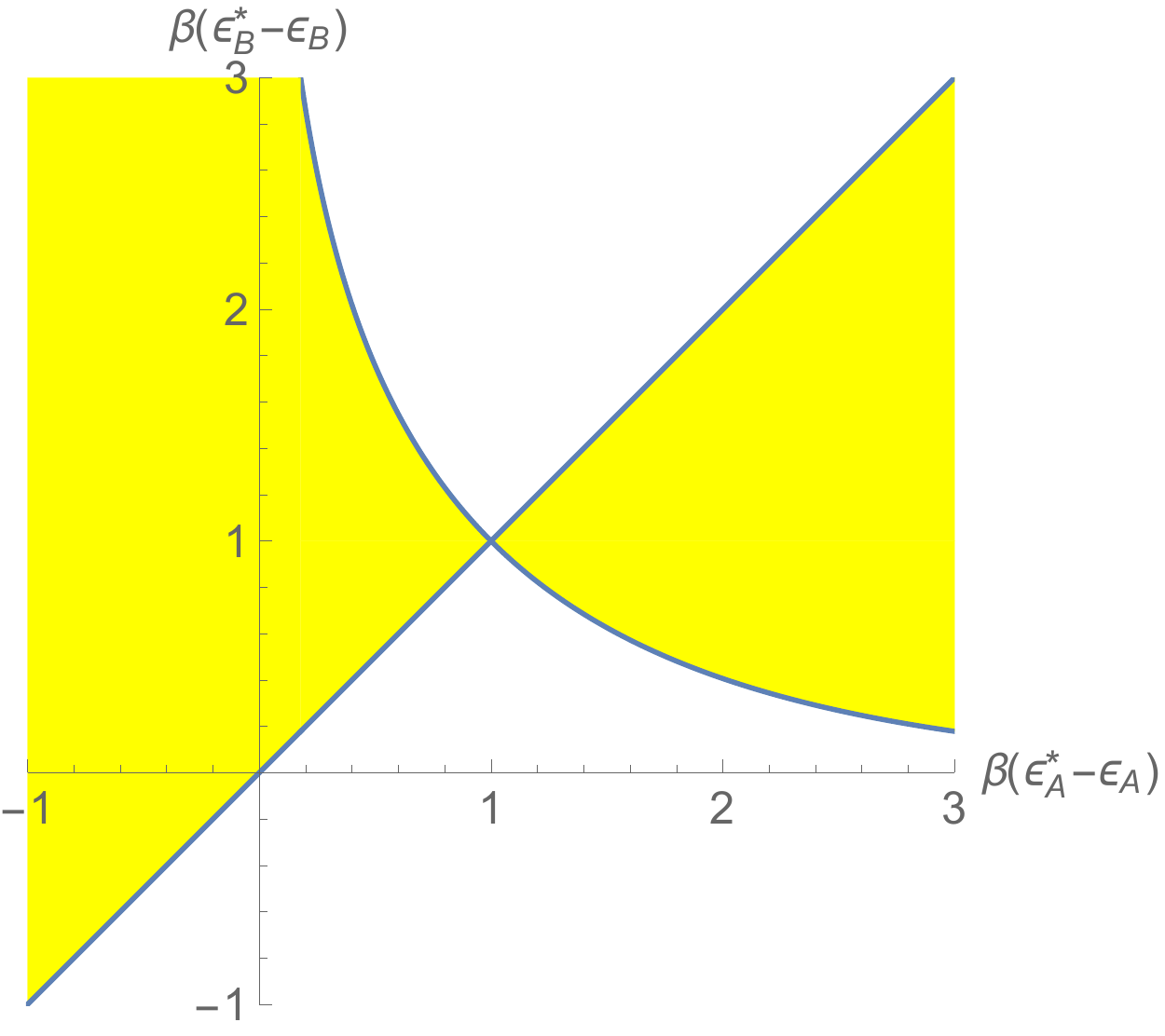}\\
  \caption{Diagram of regimes for two catalytic spheres in terms of energies $\tilde{\varepsilon}_{A}$ and $\tilde{\varepsilon}_{B}$, according to eqn (\ref{eq:Force_droplets_compact}).  Yellow marks the region where interaction force is repulsive, in other areas it is attractive. The plot is constructed for $D_{A} = D_{B}$; the only modification required in the case $D_{A} \neq D_{B}$ is change of scales along axes.}\label{fig:Where_Attraction}
\end{figure}

The interaction force (\ref{eq:Force_droplets_compact_1}) between catalytic colloids can be either attractive or repulsive.  To see this, consider the following simple model.  Imagine that catalysis takes place in a narrow layer of thickness $d$ around the catalyst surface, then
\begin{subequations} \begin{align} k_{\rightarrow} = \frac{1}{\tau} e^{\beta \left(\varepsilon_{A} - \varepsilon^{\dag} \right)} \;  \ \  k_{\leftarrow} = \frac{1}{\tau} e^{\beta \left(\varepsilon_{B} - \varepsilon^{\dag} \right)} \ , \end{align}
where $1/\tau$ is the attempt rate, $\beta = 1/k_{B}T$, while $\varepsilon_{A}$ and $\varepsilon_{B}$ are the bulk free energies of $A$ and $B$, and $\varepsilon^{\dag}$ is the free energy of the transition state of the catalytic surface reaction (for reasons of brevity, we will refer to these free energies as energies in the following). Furthermore, if energies of $A$ and $B$ molecules inside surface layer of a colloid $\varepsilon_{A}^{\ast}$ and $\varepsilon_{B}^{\ast}$ are different from their bulk values $\varepsilon_{A}$ and $\varepsilon_{B}$, then $\sigma^{\prime}_{A} = d \tilde{\varepsilon}_{A} e^{-\beta \tilde{\varepsilon}_{A}}$ and $\sigma^{\prime}_{B} = d \tilde{\varepsilon}_{B} e^{-\beta \tilde{\varepsilon}_{B}}$, with $ \tilde{\varepsilon}_{A} = \varepsilon_{A}^{\ast} - \varepsilon_{A}$ and $ \tilde{\varepsilon}_{B} = \varepsilon_{B}^{\ast} - \varepsilon_{B}$. In this approximation,
\begin{align}  \frac{J}{4 \pi R^2 d} & = \frac{c_{A}^{\infty}e^{\beta \varepsilon_{A}}  - c_{B}^{\infty}e^{\beta \varepsilon_{B}}}{ \frac{Rd}{D_{A}} e^{\beta  \varepsilon_{A}} + e^{\beta \varepsilon^{\dagger}}\tau  + \frac{Rd}{ D_{B}} e^{\beta  \varepsilon_{B}}} \label{eq:current_4} \\
\frac{f}{4\pi R^2} & = \frac{Jd}{4 \pi r^2} \left[ \frac{\tilde{\varepsilon}_{B}}{D_{B}} e^{-\beta \tilde{\varepsilon}_{B}} - \frac{\tilde{\varepsilon}_{A}}{D_{A}} e^{-\beta \tilde{\varepsilon}_{A}} \right] \ . \label{eq:Force_droplets_compact} \end{align} \label{eq:naive} \end{subequations}
Inspection of eqn (\ref{eq:Force_droplets_compact}) confirms that the force between catalysts can be attractive or repulsive, depending on the energies $\varepsilon_{A}$, $\varepsilon_{B}$, $\varepsilon_{A}^{\ast}$ and $\varepsilon_{B}^{\ast}$, as shown in the Fig. \ref{fig:Where_Attraction} (see also \cite{PhysRevLett.115.258301}).  For instance, if both $A$ and $B$ molecules are attracted to the surfaces of catalytic particles, $\varepsilon_{A}^{\ast} < \varepsilon_{A}$ and $\varepsilon_{B}^{\ast} < \varepsilon_{B}$, then the resulting long range interaction between catalysts is a competition: interaction with $A$ pushes each sphere away from the other, towards greater supply of $A$, but interaction with $B$ pulls catalysts towards one another, towards where new $B$ is produced.  Therefore, overall attraction between spheres occurs if $\varepsilon_{B}^{\ast} - \varepsilon_{B} < \varepsilon_{A}^{\ast} - \varepsilon_{A}$, and overall repulsion takes place in the opposite case.

In equations (\ref{eq:naive}), we expressed phenomenological quantities $\sigma^{\prime}_{A}$ and $\sigma^{\prime}_{B}$, as well as rate constants $k_{\rightarrow}$ and $k_{\leftarrow}$ in terms of energies such as $\varepsilon_{B}^{\ast}$, $\varepsilon_{B}$, $\varepsilon_{A}^{\ast}$, $\varepsilon_{A}$; these mechanical quantities are easy to imagine for a theorist, but virtually impossible to measure.  Furthermore, we consider only the force acting on catalytic colloidal particles, which is, in principle, measurable in an optical tweezers experiment, but we do not consider their motion under this force.  Translating force into velocity requires the knowledge of mobility, and simple minded assumption of Stokes friction is known to be only qualitatively and not quantitatively correct.  More systematic phenomenological treatments \cite{PhysRevLett.108.038303, PhysRevLett.112.068301} operate with directly measurable surface tensions, Onsager coefficients, and other phenomenological parameters.  We present the somewhat more naive approach based on equations (\ref{eq:naive}) only because of its simplicity and pedagogical value.

Our consideration so far was restricted to spherically symmetric catalytic particles.  This idealization is perhaps rarely realized.  A catalytic particle without spherical symmetry creates non-isotropic concentration fields of reagents, which can result in auto-diffusiophoretic motion of the catalyst \cite{Golestanian_2005_PhysRevLett.94.220801, Golestanian_2007, OsmoticMotor_PhysRevLett.100.158303, OsmoticMotor_Comment1_PhysRevLett.102.159801, OsmoticMotor_reply1_PhysRevLett.102.159802, OsmoticMotor_Comment2_PhysRevLett.103.079801, OsmoticMotor_Reply2_PhysRevLett.103.079802, Brady2011, Palacci2013, PhysRevLett.115.258301, Buttioni_PhysRevLett.110.238301, Popescu2016, Brady2016}.  Such self-diffusiophoretic particles are of considerable current interest and represent an important example of the so-called active swimmers \cite{Bocquet2012, Palacci2013, Buttioni_PhysRevLett.110.238301, Popescu2016}.  Speaking about concentration field mediated interactions between catalysts, we should think of multipole expansion of the concentration fields (see also \cite{Brady2016}).  Then, exactly as in the familiar electrostatics context, the dominant long range contribution is that from a monopole, $\sim 1/r^2$, which is what we considered above, while dipole (like for Janus particles), quadrupole, and higher order multipoles are important for the near field.  Thus, we will continue working in the monopole approximation which is only justified when distance between catalysts is large.  Accordingly, we do not consider self-diffusiophresis, simply because it was already studied in detail \cite{Golestanian_2005_PhysRevLett.94.220801, Golestanian_2007, PhysRevLett.108.038303, Buttioni_PhysRevLett.110.238301, PhysRevLett.112.068301, PhysRevLett.115.258301, PhysRevE.89.062316, Saha_2019}.

\section{A crowd of catalysts}\label{sec:many_catalysts}

We now turn from considering the force between two catalytic particles to the case when there are many catalysts.  Since catalysts reduce free energy barriers, thus paving the way for the system to approach chemical equilibrium, in order to maintain the system away from equilibrium in the presence of a finite concentration of catalysts, it is necessary (though maybe not sufficient) to confine the catalysts to a region of space that is surrounded by a ``bath'' in which non-equilibrium concentrations of $A$ and $B$ molecules are enforced and maintained from outside. Our goal now is to explore implications of such boundary conditions.

Consider a crowd of catalytic particles, with density of $\rho(\mathbf{r})$ catalysts per unit volume.  On the mean field level, overall behavior should be described by the volume fraction of catalytic centers in space, $\phi(\mathbf{r}) = v \rho(\mathbf{r})$, where $v$ (exactly as before) is the volume of the region in which catalysis takes place on the surface of one catalytic particle.  Now, let $c_{A}(\mathbf{r})$ and $c_{B}(\mathbf{r})$ be the concentration fields of ``fuel'' and ``exhaust'' molecules $A$ and $B$, coarse grained over distances large compared to the typical distance between catalysts, $\ell$ ($\ell^{-3} \sim \rho$).  Then mean field equations for concentrations of $A$ and $B$ read (the upper dot indicates time derivative)
\begin{equation}\begin{split} \dot{c}_{A} (\mathbf{r}) & = D_{A} \nabla^2 c_{A}(\mathbf{r})  - \phi(\mathbf{r}) \left[k_{\rightarrow} c_{A}(\mathbf{r}) - k_{\leftarrow} c_{B}(\mathbf{r}) \right] \\  \dot{c}_{B} (\mathbf{r}) & = D_{B} \nabla^2 c_{B} (\mathbf{r})+ \phi(\mathbf{r}) \left[k_{\rightarrow} c_{A}(\mathbf{r}) - k_{\leftarrow} c_{B}(\mathbf{r}) \right]\end{split} \label{eq:diffusion_eq_for_two} \end{equation}

To analyze these equations, we introduce the ``field of chemical imbalance''
\begin{equation} \psi(\mathbf{r}) \equiv k_{\rightarrow}  c_{A}(\mathbf{r}) - k_{\leftarrow} c_{B}(\mathbf{r})  \ . \label{eq:imbalance}\end{equation}
The meaning of $\psi (\mathbf{r})$ field (\ref{eq:imbalance}) is clarified by noticing that, up to a constant factor $1/v$, $\psi(\mathbf{r})$ is equal to $J(\mathbf{r})$ -- the rate of chemical reaction (\ref{eq:current_general}) in the vicinity of point $\mathbf{r}$, coarse grained over the scale $\ell$ in the same way as concentrations and $\psi$.   Therefore, according to eqn (\ref{eq:Force_droplets_compact_1}),  it also follows that $\psi(\mathbf{r})$ determines the strength of interactions between catalytic colloids around $\mathbf{r}$.

In steady state, time derivatives vanish and, combining Eqs. (\ref{eq:diffusion_eq_for_two}) with proper weights, we find that $\psi(\mathbf{r})$ satisfies
\begin{equation} \nabla^2 \psi(\mathbf{r})  - \xi^{-2}(\mathbf{r})\psi(\mathbf{r})  = 0 \ , \label{eq:screening}\end{equation}
where
\begin{equation} \xi^{-2}(\mathbf{r})  =  \left[ \frac{k_{\rightarrow}}{D_{A}} + \frac{k_{\leftarrow}}{D_{B}}\right] \phi(\mathbf{r}) \ . \label{eq:Penetration_Depth} \end{equation}

At first glance, eqn (\ref{eq:screening}) is identical to the celebrated Debye-H\"{u}ckel equation  \cite{Debye_Screening_1923} for the electrostatic potential around a point charge in an ionic solution, implying that $\xi(\mathbf{r})$ can be interpreted as the screening length (see equation (8) in reference \cite{PhysRevE.89.062316}).  Upon further inspection one notices significant differences.  First and foremost, electrostatic potential is defined up to an additive constant, while field of chemical imbalance does not have this gauge freedom, because $\psi = 0$ is the special state of chemical equilibrium.   Because of that, the boundary conditions on our chemical imbalance field $\psi(\mathbf{r})$ are quite different from those on the potential in a typical electrostatics problem: while the  electric potential diverges at the point charge and decays to a constant, usually identified as zero, at large distance from it,  $\psi$ is maintained at some fixed non-equilibrium value away from the catalysts where $\phi(\mathbf{r})=0$.  Depending on the geometry, $\psi$ vanishes or reaches some lower value inside the colloid-occupied region ($\phi(\mathbf{r})\neq 0$) because catalysis always tends to reduce the degree of chemical imbalance and drive the system towards equilibrium at which detailed balance is obeyed and $\psi\mathbf(r)=0$.  This effect has not been noticed in previous works, in which the reverse reaction (the second term in eqn (\ref{eq:imbalance})) that drives the system to equilibrium, was omitted.

\begin{figure}
\centering
\includegraphics[width=0.45\textwidth]{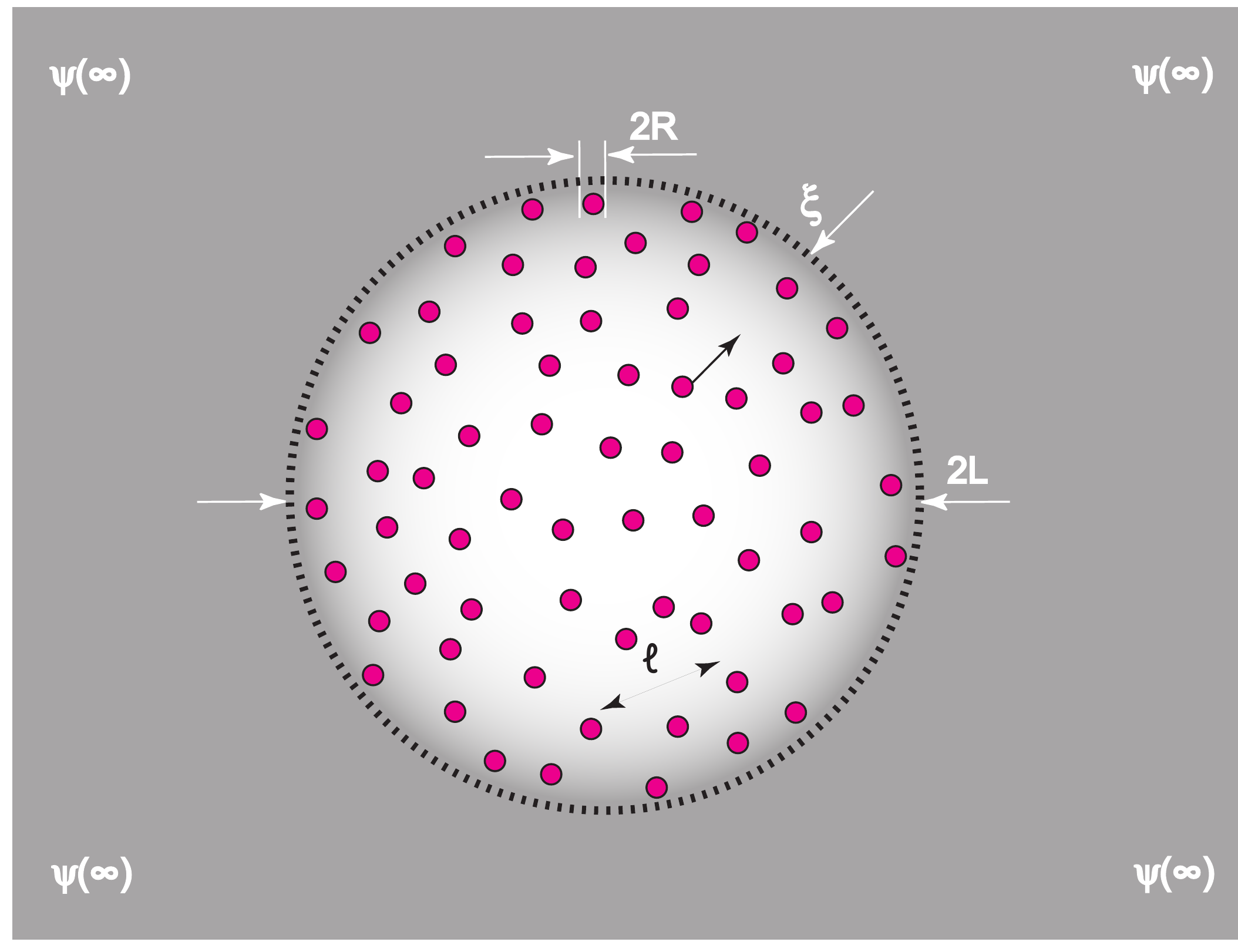}\\
\includegraphics[width=0.45\textwidth]{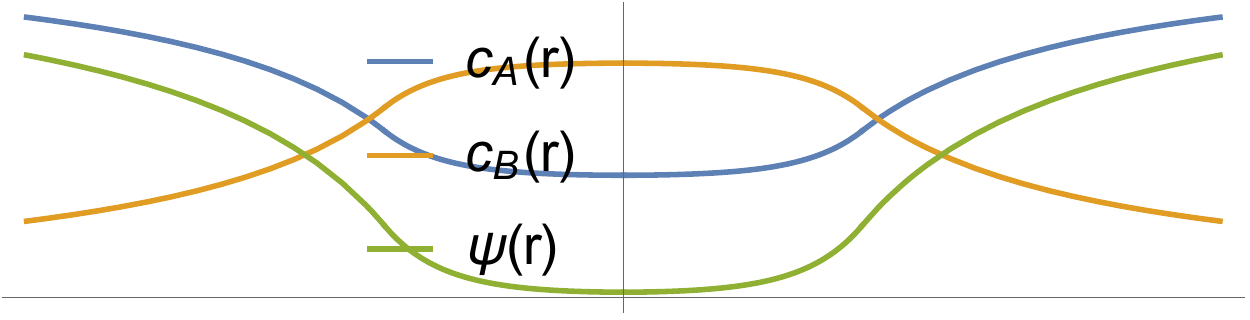}
 \caption{Catalytic particles, of diameter $2R$ each, are distributed in an osmotic bag of diameter $2L$ (shown by thick dashed circle), while substrate molecules $A$ diffuse from outside, and product molecules $B$ diffuse out to infinity.  The chemical imbalance function $k_{\rightarrow} c_{A}(r) - k_{\leftarrow} c_{B}(r) = \psi(r)$  is shown in shades of gray. Deep in the crowd of catalysts, there is no chemical imbalance between $A$ and $B$, $\psi(r) \to 0$.  For plotting, we assumed $c_{B}^{\infty} = 0$, $\xi = L/5$, $\frac{k_{\leftarrow}}{k_{\rightarrow}} =\frac{1}{2}$ and $\frac{D_{B}}{D_{A}} = 1$.}\label{fig:Penetration_Depth}
\end{figure}

In order to understand the physical meaning of the  length $\xi$ let us consider the 3D situation shown in Fig. \ref{fig:Penetration_Depth}: catalytic colloids are confined inside a spherical osmotic bag of radius $L$ which is permeable to solute molecules $A$ and $B$ but not to colloids. Substrate molecules $A$ are delivered by diffusion from infinity, and product molecules $B$ are also absorbed at infinity such that their concentrations at infinity are fixed at some non-equilibrium values $c_{A}^{\infty}$ and $c_{B}^{\infty}$, respectively (note that since chemical reactions take place only inside the bag, these bulk concentrations can be arbitrarily far from equilibrium). As shown in Fig. \ref{fig:Penetration_Depth}, the chemical imbalance field $\psi(r)$ penetrates up to a {\it penetration depth} $\xi$ into the catalysts-occupied domain. Deeper into the bulk of the catalysts-occupied region $\psi(r)\rightarrow 0$, the concentrations of $A$ and $B$ approach equilibrium values.   More specifically, assuming the concentrations to be $c_{A}^{\infty}$ and $c_{B}^{\infty}$ at infinity, concentration profiles are expressed in terms of chemical imbalance function $\psi(r)$
\begin{subequations} \begin{align} c_{A}(r) & = c_{A}^{\infty} + \frac{ \psi(r) - \psi(\infty)}{D_A \left(\frac{k_{\rightarrow}}{D_{A}} + \frac{k_{\leftarrow}}{D_{B}} \right)} \\ c_{B}(r) & = c_{B}^{\infty} - \frac{ \psi(r) - \psi(\infty)}{D_B \left(\frac{k_{\rightarrow}}{D_{A}} + \frac{k_{\leftarrow}}{D_{B}} \right)}  \ , \end{align}
while $\psi(r)$ itself is found for this spherical geometry, based on eqn (\ref{eq:screening}), along with boundary conditions of continuous function and its derivative and no singularity at the origin:
\begin{align} \frac{\psi(r)}{ \psi(\infty)} =  \left\{ \begin{array}{lcr} 1 - \frac{L}{r} + \frac{\xi}{r} \tanh \frac{L}{\xi}  & \mathrm{ at} & r>L \\  \frac{\xi}{r} \frac{\sinh r/\xi}{\cosh L/\xi} & \mathrm{at} & r < L \end{array} \right. \label{eq:solution_for_psi} \end{align} \label{eq:solution_in_3D} \end{subequations}
These results are plotted, for specific values of parameters, in Fig. \ref{fig:Penetration_Depth}.  As expected, the current $J$ vanishes inside the crowd of catalysts along with $\psi$, and the forces between colloids vanish as well. These forces (attractive or repulsive) will be significant only inside the boundary layer of thickness $\xi$. If the size of the osmotic bag $L$ is smaller or comparable to the penetration depth $\xi$, depending on the sign of the force in eqn (\ref{eq:Force_droplets_compact}), catalysts will attract one another and form an aggregate (which could be smaller than the available volume of the osmotic bag) or repel each other and form a Wigner crystal \cite{Wigner_PhysRev.46.1002} (occupying the whole accessible volume).

\begin{figure}
\centering
\includegraphics[width=0.45\textwidth]{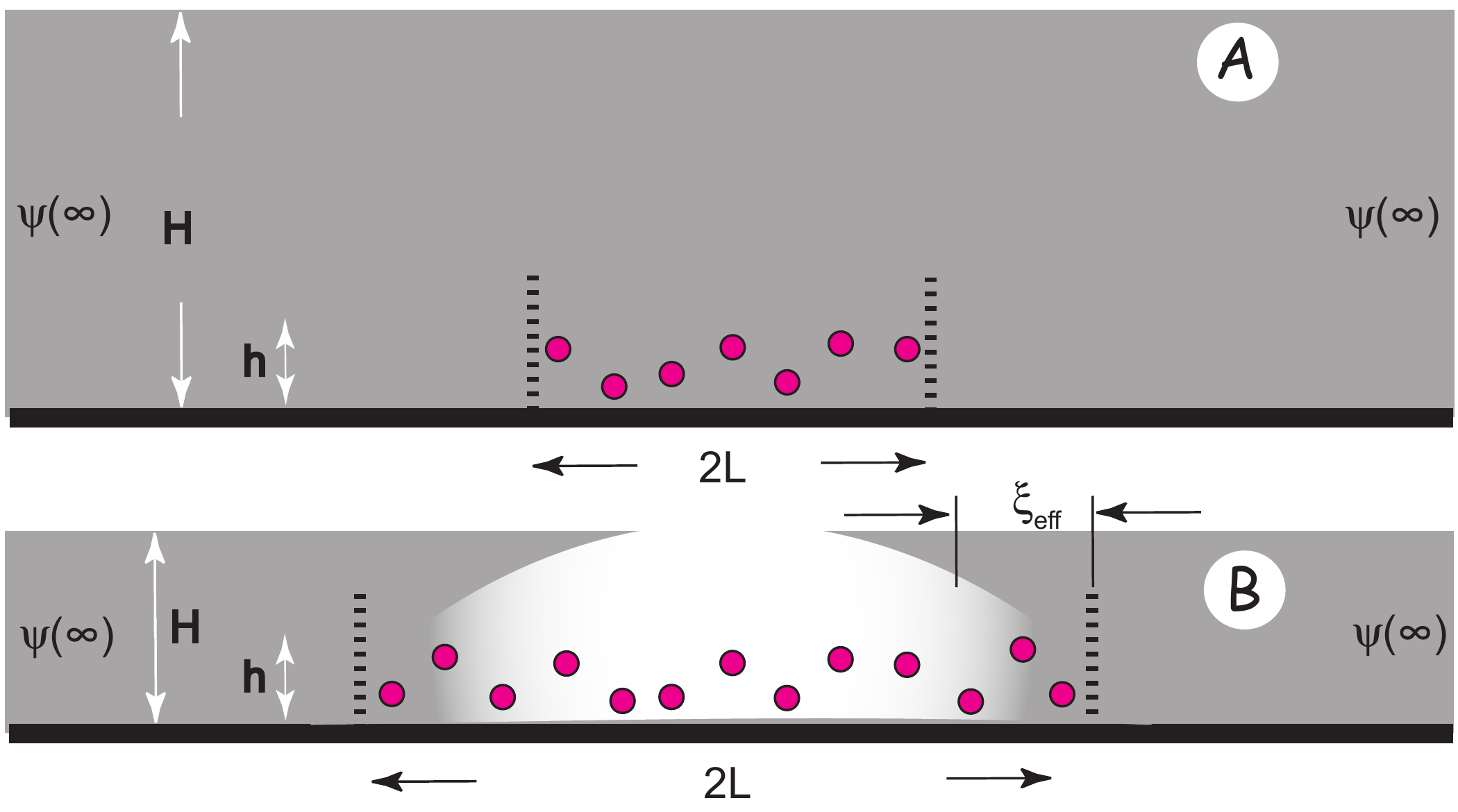}\\
\caption{Catalytic colloids are located within a gravitational height $h$ from the bottom of the container of depth $H$.  As before, the chemical imbalance field is approximately shown by shades of gray.  Barriers or osmotic bag are shown by thick dashed lines. In case of a deep container (figure A), diffusion in 3D is sufficient to supply fuel and remove exhaust for all colloids, and chemical imbalance field $\psi$ penetrates to all colloids.  In case of a shallow container (figure B), the middle part of the ``pancake'' of colloids is not accessible to chemical imbalance field. }\label{fig:Quasi_2D}
\end{figure}

A more subtle and experimentally relevant case is a quasi-2D system, where the container has a finite depth $H$, while catalytic colloids are confined due to gravity within a short distance $h$, sometimes called gravitational height, from the bottom (or from the top if they float) of a container, as shown in a cartoon, Fig. \ref{fig:Quasi_2D}.  Boundary conditions, in addition to fixed value of $\psi(\infty)$ at $r \to \infty$, require zero normal (vertical) flux of either $A$ or $B$ particles, thus vanishing normal derivative of $\psi$ on both top and bottom surfaces (we note in passing that this does not have simple electrostatic analogy).  The situation, as it turns out, depends sensitively on the relations between several relevant length scales.

If the depth of the container is infinite or very large, as in Fig. \ref{fig:Quasi_2D}A, then although colloidal spheres are confined in 2D by barriers  or osmotic bag to the interior of a circle of radius $L$, the fuel $A$ and exhaust $B$ molecules are diffusing in 3D.  Formally, in this case, equation (\ref{eq:screening}) gets reduced simply to $\nabla^2 \psi = 0$ everywhere except a very thin pancake-shaped region of thickness $h$ and radius $L$, and finite penetration depth $\xi$ (\ref{eq:Penetration_Depth}) exists only inside the pancake.  If $h$ is very small and $\xi \gg h$, then delivery of $A$ and removal of $B$ by diffusion in 3D is unhindered and reaches every point of the pancake from the top.  Therefore, the force of interaction between catalytic colloids still obeys the $1/r^2$ law everywhere inside the pancake, unlike the 3D case where colloids in the bulk of the confined region essentially do not interact.  In the attractive case, we expect catalysts in 2D to form a large aggregate whose growth may be stopped only when its diameter becomes comparable to $H$.  For repulsive forces we expect formation of a 2D Wigner crystal whose size is not limited by $\xi$ and is controlled by the confining boundaries or an osmotic bag only.

Consider now the opposite limit of a shallow container, as in Fig. \ref{fig:Quasi_2D}B. Clearly, on horizontal length scales larger than $H$ the problem becomes essentially two dimensional. Averaging eqn (\ref{eq:screening}) along the vertical direction, we obtain 2D equation of the same form, except with effective penetration depth given by $\xi^{-2}_{\mathrm{eff}} = \xi^{-2} h/H$.  The solution of the corresponding 2D problem is qualitatively similar to Eqs. (\ref{eq:solution_in_3D}), which means that the field of chemical imbalance penetrates into the pancake of catalysts only to horizontal distance about $\xi_{\mathrm{eff}} = \xi \left( H/h \right)^{1/2}$.  Closer to the middle of the pancake the local imbalance field is reduced by catalysis and long-range interactions between colloids are suppressed.  In this case we expect attractive catalysts to assemble in 2D aggregates of size no larger than $\sim \xi_{\mathrm{eff}}$. Repulsive catalysts will form 2D Wigner crystals only for $L\leq \xi_{\mathrm{eff}}$.

This conclusion is reminiscent of the fact that ``live'' colloids in the experiments by Pallaci et al \cite{Palacci2013}, Buttioni et al \cite{Buttioni_PhysRevLett.110.238301}, and Theurkauff et al \cite{Bocquet2012} formed 2D aggregates of limited size that did not grow further.  These colloids were not spherically-symmetric and exhibited self-diffusiophoretic swimming.  Moreover, they were shown \cite{Palacci2013} to form the so-called ``living crystals'', a finding which was interpreted as an experimental confirmation of the theoretically predicted activity-driven condensation  \cite{PhysRevLett.108.235702, PhysRevLett.110.055701, PhysRevLett.110.238301, PhysRevLett.111.145702, Bialk__2013, C3SM52813H, C3SM52469H, Wysocki_2014, PhysRevLett.112.218304, Cates_phi_4_2014}.  We speculate that the limited size of the aggregates could be due to  the finite penetration depth of the chemical imbalance field (this effect does not require catalysts to be spherically symmetric and is expected to take place even for self-diffusiophoretically driven swimmers).

\section{Conclusion}

To summarize, we have presented a very simple schematic theory demonstrating that spherical colloids capable of catalyzing a reversible chemical reaction between solutes, experience a peculiar interaction which exists only as long as the concentrations of the solutes are maintained out of equilibrium  by constant supply of high free energy ``fuel'' and removal of ``exhaust'' molecules at the boundaries of the colloid solution. The long range ($1/r$) interaction between colloids is driven by the chemical imbalance field that measures the deviation from chemical equilibrium. We have shown that, despite the apparent similarity of the underlying equations, the origin of this effect is very different from the Debye-Huckel electrolyte polarization mechanism in electrostatics: catalytic activity drives the concentrations of solute molecules towards their equilibrium values and therefore reduces the chemical imbalance that controls the strength of the diffusiophoretic interaction between the colloids.

We demonstrated that the combination of boundary conditions and finite penetration depth has a profound effect on the interaction between catalytic colloids. Thus, in a realistic 3D geometry of a colloid solution enclosed in an osmotic bag (permeable to solute molecules but not to colloids) and surrounded by a ``bath'' that fixes the concentrations of solutes at some arbitrary values, non-equilibrium concentrations of solutes can be maintained in steady state only within a penetration depth from the boundary, and therefore interactions between colloids vanish in the bulk of the colloid solution.   These results remain valid even in the limit of vanishing reverse reaction rate, when they boil down to a simple statement that fuel molecules $A$ cannot penetrate into the bulk of the catalyst crowd as they get transformed into $B$ (the field $\psi(r)$ is very small at small $r$ according to eqn (\ref{eq:solution_for_psi}), meaning in eqn (\ref{eq:imbalance}) that $k_{\leftarrow} \to 0$ and $c_{A}(r) \to 0$); we thank the Referee for bringing our attention to this point. The effects of finite penetration depth can be overcome in quasi-2D geometry (with colloids confined to a surface and solute molecules free to move in 3D) where unscreened $1/r$ attractions between colloids can lead to macroscopic aggregates or to Wigner crystals, depending on the sign of the diffusiophoretic interaction between colloids. In the attractive case, we predict that finite 2D clusters of colloids will form if the depth of the 3D container is finite and that the   diameter of these clusters will be proportional to the effective correlation length that increases as the square root of the depth $H$. In the repulsive case one expects Wigner crystals to form if the separtion between the barriers that confine the colloids in 2D is smaller than the effective penetration length. These theoretical predictions await experimental verification.

\acknowledgements
We would like to thank Alexandra Zidovska and Paul Chaikin for valuable discussions. We are also indebted to Ramin Golestanian for helpful comments and suggestions, and thank Siegfried Dietrich and Mihail Popescu for useful correspondence. YR's work was supported by grants 178/16 from the Israel Science Foundation and 1902/12 from the Israeli Centers for Research Excellence program of the Planning and Budgeting Committee. YR would like to acknowledge the hospitality of the Center for Soft Matter Research of New York University where part of this work was done.  AYG's research is supported in part by the MRSEC Program of the National Science Foundation under Award DMR-1420073.  This research was supported in part by the National Science Foundation under Grant No. NSF PHY-1748958 and by the National Institutes of Health under Grant No. R25GM067110.


\begin{thebibliography}{10}

\bibitem{Motors_Review_RevModPhys.69.1269}
Frank J\"ulicher, Armand Ajdari, and Jacques Prost.
\newblock Modeling molecular motors.
\newblock {\em Rev. Mod. Phys.}, 69:1269--1282, 1997.

\bibitem{ColloidalSwimmers_EBBENS201614}
S.J. Ebbens.
\newblock Active colloids: Progress and challenges towards realising autonomous
  applications.
\newblock {\em Current Opinion in Colloid \& Interface Science}, 21:14 -- 23,
  2016.

\bibitem{Anderson1989}
John~L. Anderson.
\newblock Colloid transport by interfacial forces.
\newblock {\em Ann. Rev. Fluid Mech.}, 21:61--69, 1989.

\bibitem{Brady2011}
John~F. Brady.
\newblock Particle motion driven by solute gradients with application to
  autonomous motion: continuum and colloidal perspectives.
\newblock {\em J. Fluid. Mech.}, 667:216--259, 2011.

\bibitem{Paustian2015}
Joel~S. Paustian, Craig~D. Angulo, Rodrigo Nery-Azevedo, Nan Shi, Amr~I.
  Abdel-Fattah, and Todd~M. Squires.
\newblock Direct measurements of colloidal solvophoresis under imposed solvent
  and solute gradients.
\newblock {\em Langmuir}, 31(15):4402--4410, 2015.

\bibitem{Sear2017}
Richard~P. Sear and Patrick~B. Warren.
\newblock Diffusiophoresis in nonadsorbing polymer solutions: The
  asakura-oosawa model and stratification in drying films.
\newblock {\em Phys. Rev. E}, 96:062602, 2015.

\bibitem{Parry2014}
Bradley~R. Parry, Ivan~V. Surovtsev, Matthew~T. Cabeen, Corey~S. O'Hern,
  Eric~R. Dufresne, and Christine Jacobs-Wagner.
\newblock The bacterial cytoplasm has glass-like properties and is fluidized by
  metabolic activity.
\newblock {\em Cell}, 156:183–194, 2014.

\bibitem{Sear2019}
Richard~P. Sear.
\newblock Diffusiophoresis in cells: a general non-equilibrium, non-motor
  mechanism for the metabolism-dependent transport of particles in cells.
\newblock {\em Phys. Rev. Lett.}, 122:128101, 2019.

\bibitem{Vanag2009}
Vladimir~K. Vanag and Irving~R. Epstein.
\newblock Cross-diffusion and pattern formation in reaction–diffusion
  systems.
\newblock {\em Phys. Chem. Chem. Phys.}, 11:897–912, 2009.

\bibitem{Zhao2018}
Xi~Zhao, Henri Palacci, Vinita Yadav, Michelle~M. Spiering, Michael~K. Gilson,
  and Hess~Henry Butler, Peter~J., Stephen~J. Benkovic, and Ayusman Sen.
\newblock Substrate-driven chemotactic assembly in an enzyme cascade.
\newblock {\em Nature Chemistry}, 10:311–317, 2018.

\bibitem{PhysRevLett.108.038303}
Ramin Golestanian.
\newblock Collective behavior of thermally active colloids.
\newblock {\em Phys. Rev. Lett.}, 108:038303, 2012.

\bibitem{PhysRevLett.112.068301}
Rodrigo Soto and Ramin Golestanian.
\newblock Self-assembly of catalytically active colloidal molecules: Tailoring
  activity through surface chemistry.
\newblock {\em Phys. Rev. Lett.}, 112:068301, 2014.

\bibitem{PhysRevE.89.062316}
Suropriya Saha, Ramin Golestanian, and Sriram Ramaswamy.
\newblock Clusters, asters, and collective oscillations in chemotactic
  colloids.
\newblock {\em Phys. Rev. E}, 89:062316, 2014.

\bibitem{PhysRevE.91.052304}
Rodrigo Soto and Ramin Golestanian.
\newblock Self-assembly of active colloidal molecules with dynamic function.
\newblock {\em Phys. Rev. E}, 91:052304, 2015.

\bibitem{PhysRevLett.123.018101}
Jaime Agudo-Canalejo and Ramin Golestanian.
\newblock Active phase separation in mixtures of chemically interacting
  particles.
\newblock {\em Phys. Rev. Lett.}, 123:018101, 2019.

\bibitem{Saha_2019}
Suropriya Saha, Sriram Ramaswamy, and Ramin Golestanian.
\newblock Pairing, waltzing and scattering of chemotactic active colloids.
\newblock {\em New Journal of Physics}, 21(6):063006, 2019.

\bibitem{Golestanian_Review_2019}
Ramin Golestanian.
\newblock Phoretic active matter.
\newblock arXiv:1909.03747, 2019.

\bibitem{Nasouri_Golestanian_2020}
Babak Nasouri and Ramin Golestanian.
\newblock Exact phoretic interaction of two chemically-active particles.
\newblock arXiv:2001.07576, 2020.

\bibitem{PhysRevLett.115.258301}
Benno Liebchen, Davide Marenduzzo, Ignacio Pagonabarraga, and Michael~E. Cates.
\newblock Clustering and pattern formation in chemorepulsive active colloids.
\newblock {\em Phys. Rev. Lett.}, 115:258301, 2015.

\bibitem{PhysRevE.81.046311}
Christel Hohenegger and Michael~J. Shelley.
\newblock Stability of active suspensions.
\newblock {\em Phys. Rev. E}, 81:046311, 2010.

\bibitem{Oshanin_2017}
Gleb Oshanin, Mihail~N. Popescu, and Siegfried Dietrich.
\newblock Active colloids in the context of chemical kinetics.
\newblock {\em Journal of Physics A: Mathematical and Theoretical},
  50(13):134001, 2017.

\bibitem{Debye_Screening_1923}
Von~P. Debye and E.~H\"{u}ckel.
\newblock Zur theorie der elektrolyte.
\newblock {\em Physikalische Zeitschrift}, 24:185 -- 206, 1923.

\bibitem{Wigner_PhysRev.46.1002}
Eugene Wigner.
\newblock On the interaction of electrons in metals.
\newblock {\em Phys. Rev.}, 46:1002--1011, 1934.

\bibitem{Reversible_MichaelisMenten_1}
W.~Grady Smith.
\newblock In vivo kinetics and the reversible {Michaelis-Menten} model.
\newblock {\em Journal of Chemical Education}, 69(12):981, 1992.

\bibitem{Reversible_MichaelisMenten_2}
Reversible {Michaelis-Menten} kinetics.
\newblock \begin{verbatim}http://www.bio-physics.at/wiki\end{verbatim}, 2014.

\bibitem{SupMat1}
See supplementary material at the end of this file.

\bibitem{Golestanian_2005_PhysRevLett.94.220801}
Ramin Golestanian, Tanniemola~B. Liverpool, and Armand Ajdari.
\newblock Propulsion of a molecular machine by asymmetric distribution of
  reaction products.
\newblock {\em Phys. Rev. Lett.}, 94:220801, 2005.

\bibitem{Golestanian_2007}
R.~Golestanian, T.~B. Liverpool, and A.~Ajdari.
\newblock Designing phoretic micro- and nano-swimmers.
\newblock {\em New Journal of Physics}, 9(5):126--126, 2007.

\bibitem{OsmoticMotor_PhysRevLett.100.158303}
Ubaldo~M. C\'ordova-Figueroa and John~F. Brady.
\newblock Osmotic propulsion: The osmotic motor.
\newblock {\em Phys. Rev. Lett.}, 100:158303, 2008.

\bibitem{OsmoticMotor_Comment1_PhysRevLett.102.159801}
Thomas~M. Fischer and Prajnaparamita Dhar.
\newblock Comment on ``{Osmotic Propulsion: The Osmotic Motor}''.
\newblock {\em Phys. Rev. Lett.}, 102:159801, 2009.

\bibitem{OsmoticMotor_reply1_PhysRevLett.102.159802}
Ubaldo~M. C\'ordova-Figueroa and John~F. Brady.
\newblock {C\'ordova-Figueroa} and {Brady} reply.
\newblock {\em Phys. Rev. Lett.}, 102:159802, 2009.

\bibitem{OsmoticMotor_Comment2_PhysRevLett.103.079801}
Frank J\"ulicher and Jacques Prost.
\newblock Comment on ``{Osmotic Propulsion: The Osmotic Motor}''.
\newblock {\em Phys. Rev. Lett.}, 103:079801, 2009.

\bibitem{OsmoticMotor_Reply2_PhysRevLett.103.079802}
Ubaldo~M. C\'ordova-Figueroa and John~F. Brady.
\newblock {C\'ordova-Figueroa} and {Brady} reply.
\newblock {\em Phys. Rev. Lett.}, 103:079802, 2009.

\bibitem{Palacci2013}
Jeremie Palacci, Stefano Sacanna, Asher~Preska Steinberg, David~J. Pine, and
  Paul~M. Chaikin.
\newblock Living crystals of light-activated colloidal surfers.
\newblock {\em Science}, 339(6122):936--940, 2013.

\bibitem{Buttioni_PhysRevLett.110.238301}
Ivo Buttinoni, Julian Bialk\'e, Felix K\"ummel, Hartmut L\"owen, Clemens
  Bechinger, and Thomas Speck.
\newblock Dynamical clustering and phase separation in suspensions of
  self-propelled colloidal particles.
\newblock {\em Phys. Rev. Lett.}, 110:238301, 2013.

\bibitem{Popescu2016}
Mihail~N. Popescu, William~E. Uspal, and Siegfried Dietrich.
\newblock Self-diffusiophoresis of chemically active colloids.
\newblock {\em Eur. Phys. J. Special Topics}, 225:2189–2206, 2016.

\bibitem{Brady2016}
Wen Yan and John~F. Brady.
\newblock The behavior of active diffusiophoretic suspensions: An accelerated
  laplacian dynamics study.
\newblock {\em J. Chem. Phys.}, 145:134902, 2016.

\bibitem{Bocquet2012}
I.~Theurkauff, C.~Cottin-Bizonne, J.~Palacci, C.~Ybert, and L.~Bocquet.
\newblock Dynamic clustering in active colloidal suspensions with chemical
  signaling.
\newblock {\em Phys. Rev. Lett.}, 108:268303, 2012.

\bibitem{PhysRevLett.108.235702}
Yaouen Fily and M.~Cristina Marchetti.
\newblock Athermal phase separation of self-propelled particles with no
  alignment.
\newblock {\em Phys. Rev. Lett.}, 108:235702, 2012.

\bibitem{PhysRevLett.110.055701}
Gabriel~S. Redner, Michael~F. Hagan, and Aparna Baskaran.
\newblock Structure and dynamics of a phase-separating active colloidal fluid.
\newblock {\em Phys. Rev. Lett.}, 110:055701, 2013.

\bibitem{PhysRevLett.110.238301}
Ivo Buttinoni, Julian Bialk\'e, Felix K\"ummel, Hartmut L\"owen, Clemens
  Bechinger, and Thomas Speck.
\newblock Dynamical clustering and phase separation in suspensions of
  self-propelled colloidal particles.
\newblock {\em Phys. Rev. Lett.}, 110:238301, 2013.

\bibitem{PhysRevLett.111.145702}
Joakim Stenhammar, Adriano Tiribocchi, Rosalind~J. Allen, Davide Marenduzzo,
  and Michael~E. Cates.
\newblock Continuum theory of phase separation kinetics for active brownian
  particles.
\newblock {\em Phys. Rev. Lett.}, 111:145702, 2013.

\bibitem{Bialk__2013}
Julian Bialk{\'{e}}, Hartmut Löwen, and Thomas Speck.
\newblock Microscopic theory for the phase separation of self-propelled
  repulsive disks.
\newblock {\em {EPL} (Europhysics Letters)}, 103(3):30008, 2013.

\bibitem{C3SM52813H}
Joakim Stenhammar, Davide Marenduzzo, Rosalind~J. Allen, and Michael~E. Cates.
\newblock Phase behaviour of active brownian particles: the role of
  dimensionality.
\newblock {\em Soft Matter}, 10:1489--1499, 2014.

\bibitem{C3SM52469H}
Yaouen Fily, Silke Henkes, and M.~Cristina Marchetti.
\newblock Freezing and phase separation of self-propelled disks.
\newblock {\em Soft Matter}, 10:2132--2140, 2014.

\bibitem{Wysocki_2014}
Adam Wysocki, Roland~G. Winkler, and Gerhard Gompper.
\newblock Cooperative motion of active brownian spheres in three-dimensional
  dense suspensions.
\newblock {\em {EPL} (Europhysics Letters)}, 105(4):48004, 2014.

\bibitem{PhysRevLett.112.218304}
Thomas Speck, Julian Bialk\'e, Andreas~M. Menzel, and Hartmut L\"owen.
\newblock Effective cahn-hilliard equation for the phase separation of active
  brownian particles.
\newblock {\em Phys. Rev. Lett.}, 112:218304, 2014.

\bibitem{Cates_phi_4_2014}
Raphael Wittkowski, Adriano Tiribocchi, Joakim Stenhammar, Rosalind~J. Allen,
  Davide Marenduzzo, and Michael~E. Cates.
\newblock Scalar $\varphi^4$ field theory for active-particle phase separation.
\newblock {\em Nature Communications}, 5:4351 -- 4360, 2014.

\end{thebibliography}

\appendix

\section{Supplementary Material for ``Non-Equilibrium Interaction Between Catalytic Colloids:  Boundary Conditions and Penetration Depth'': 
Perturbation theory for the two droplets problem}\label{sec:perturbation}

Suppose now there are two spherical particles, labeled here by Roman numerals $I$ and $II$, and positioned in points ${\mathbf r}_{I}$ and ${\mathbf r}_{II}$.  We need to find steady state concentration distributions $c^{(2)}_A(\mathbf{r} | {\mathbf r}_{I}, {\mathbf r}_{II})$ and $c^{(2)}_B(\mathbf{r} | {\mathbf r}_{I}, {\mathbf r}_{II})$ through all points ${\mathbf r}$ in space.  The intuition is that when particles are far apart compared at their own sizes, $\left|  {\mathbf r}_{I} - {\mathbf r}_{II} \right| \gg R$, the solution for the two particles problem should be somehow approximated by the superposition of single particle solutions.  To formalize this idea, we first of all define $\delta c$ for the two-particles and single particle concentration profiles as the quantities that vanish in infinity, according to
\begin{subequations}\begin{align} \delta c^{(2)}_{A}(\mathbf{r} | {\mathbf r}_{I}, {\mathbf r}_{II}) & = c_{A}(\mathbf{r} | {\mathbf r}_{I}, {\mathbf r}_{II}) - c_{A} , \\ \delta c^{(1)}_{A}(\mathbf{r} | {\mathbf r}_{{I}}) & = c_{A}(\mathbf{r} | {\mathbf r}_{{I} })  - c_{A} , \\ \delta c^{(1)}_{A}(\mathbf{r} | {\mathbf r}_{{II}}) & = c_{A}(\mathbf{r} | {\mathbf r}_{{II} })  - c_{A} \ , \end{align} \end{subequations}
with similar notation also for $B$.  Upper indices $(1)$ or $(2)$ indicate solution for one or two particles, respectively.  And now we define $\delta \delta c$ encapsulating the idea of superposition; it is actually the difference between exact solution and superposition:
%
%
\begin{subequations} \begin{align} \begin{split} \delta c^{(2)}_A(\mathbf{r} | {\mathbf r}_{I}, {\mathbf r}_{II}) & =  \delta c^{(1)}_A(\mathbf{r} | {\mathbf r}_{{I}}) + \delta c^{(1)}_A(\mathbf{r} | {\mathbf r}_{{II}}) \\ & + \delta \delta c^{(2)}_A(\mathbf{r} | {\mathbf r}_{I}, {\mathbf r}_{II}) \end{split} \\ \begin{split} \delta c^{(2)}_B(\mathbf{r} | {\mathbf r}_{I}, {\mathbf r}_{II}) & =  \delta c^{(1)}_B(\mathbf{r} | {\mathbf r}_{{I}}) + \delta c^{(1)}_B(\mathbf{r} | {\mathbf r}_{{II}}) \\ & + \delta \delta c^{(2)}_B(\mathbf{r} | {\mathbf r}_{I}, {\mathbf r}_{II}) \end{split}\end{align}\end{subequations}
For now, this is nothing more than just a notation, the definition of $\delta \delta c$; our goal is now to estimate these functions and to claim that they can be neglected.  To do so, we derive, rigorously, equations and boundary conditions satisfied by these functions; we will write all of them in a standard way, with left hand side in every case representing a linear function of $\delta \delta c$, and right hand side representing inhomogeneity:
\begin{enumerate}
\begin{subequations}
\item Diffusion equations for $A$ and for $B$:
\begin{align}
\nabla^2 \delta \delta c^{(2)}_{A}(\mathbf{r} | {\mathbf r}_{I}, {\mathbf r}_{II}) & = 0 \ , \\
\nabla^2 \delta \delta c^{(2)}_{B}(\mathbf{r} | {\mathbf r}_{I}, {\mathbf r}_{II}) & = 0 \ . \label{eq:diffusion_eq}
\end{align}

\item Vanishing in infinity:
\begin{align}
\left. \delta \delta c^{(2)}_{A}(\mathbf{r} | {\mathbf r}_{I}, {\mathbf r}_{II})\right|_{r \to \infty} & = 0 \ , \\
\left. \delta \delta c^{(2)}_{B}(\mathbf{r} | {\mathbf r}_{I}, {\mathbf r}_{II})\right|_{r \to \infty} & = 0 \label{eq:venishing_infty}
\end{align}

\item Diffusion influx of $A$ continuously connects to chemical transformation flux.  For notational simplicity we introduce here rate catalytical rate constants $k_{AB} = (1/\tau) e^{\beta \left( \varepsilon_{A} - \varepsilon^{\dagger} \right)}$ and $k_{BA} = (1/\tau) e^{\beta \left( \varepsilon_{B} - \varepsilon^{\dagger} \right)}$.  This flux continuity condition must be satisfied in every point on the surface of both particles; for brevity, we write this condition only for one particle, particle ${II}$, where we specify an arbitrary point $\mathbf{r} = \mathbf{r}_{II} + \mathbf{R}$ by a vector $\mathbf{R}$ with absolute value $R$:
\begin{align}\begin{split}
& \left. D_A \nabla_{\perp} \delta \delta c^{(2)}_A(\mathbf{r} | {\mathbf r}_{I}, {\mathbf r}_{II})\right|_{\mathbf{r} = \mathbf{r}_{II} + \mathbf{R}} -  \\ &  \ \ \ \ \ \ \ \ \ \ \ \ \ \ \left. -k_{AB} \delta \delta c^{(2)}_A(\mathbf{r} | {\mathbf r}_{I}, {\mathbf r}_{II})\right|_{\mathbf{r} = \mathbf{r}_{II} + \mathbf{R}} +  \\ &  \ \ \ \ \ \ \ \ \ \ \ \ \ \ \left. +k_{BA} \delta \delta c^{(2)}_B(\mathbf{r} | {\mathbf r}_{I}, {\mathbf r}_{II})\right|_{\mathbf{r} = \mathbf{r}_{II} + \mathbf{R}} =  \\ &  = \left. - D_A \nabla_{\perp}  \delta c^{(1)}_A(\mathbf{r} |  {\mathbf r}_{I})\right|_{\mathbf{r} = \mathbf{r}_{II} + \mathbf{R} } + \\ &  \ \ \ \ \ \ \ \ \ \ \ \ \ \ \left. k_{AB} \delta  c^{(1)}_A(\mathbf{r} | {\mathbf r}_{I})\right|_{\mathbf{r} = \mathbf{r}_{II} + \mathbf{R}} +  \\ &  \ \ \ \ \ \ \ \ \ \ \ \ \ \ \left. -k_{BA} \delta c^{(1)}_B(\mathbf{r} | {\mathbf r}_{I})\right|_{\mathbf{r} = \mathbf{r}_{II} + \mathbf{R}}
\end{split}\label{eq:flux_continuity_A_on_II} \end{align}
There is also a similar condition for particle ${I}$ which we do not write for brevity.

\item Similarly to the above, diffusion outflux of $B$ continuously connects to chemical transformation flux.  Again for particle ${II}$ we have:
\begin{align}\begin{split}
& \left. D_B \nabla_{\perp} \delta \delta c^{(2)}_B(\mathbf{r} | {\mathbf r}_{I}, {\mathbf r}_{II})\right|_{\mathbf{r} = \mathbf{r}_{II} + \mathbf{R}} -  \\ &  \ \ \ \ \ \ \ \ \ \ \ \ \ \ \left. k_{AB} \delta \delta c^{(2)}_A(\mathbf{r} | {\mathbf r}_{I}, {\mathbf r}_{II})\right|_{\mathbf{r} = \mathbf{r}_{II} + \mathbf{R}} +  \\ &  \ \ \ \ \ \ \ \ \ \ \ \ \ \ \left. -k_{BA} \delta \delta c^{(2)}_B(\mathbf{r} | {\mathbf r}_{I}, {\mathbf r}_{II})\right|_{\mathbf{r} = \mathbf{r}_{II} + \mathbf{R}} =  \\ &  = \left. - D_B \nabla_{\perp}  \delta c^{(1)}_B(\mathbf{r} |  {\mathbf r}_{I})\right|_{\mathbf{r} = \mathbf{r}_{II} + \mathbf{R} } + \\ &  \ \ \ \ \ \ \ \ \ \ \ \ \ \ \left. k_{AB} \delta  c^{(1)}_A(\mathbf{r} | {\mathbf r}_{I})\right|_{\mathbf{r} = \mathbf{r}_{II} + \mathbf{R}} +  \\ &  \ \ \ \ \ \ \ \ \ \ \ \ \ \ \left. -k_{BA} \delta c^{(1)}_B(\mathbf{r} | {\mathbf r}_{I})\right|_{\mathbf{r} = \mathbf{r}_{II} + \mathbf{R}}
\end{split}\label{eq:flux_continuity_B_on_II} \end{align}

\label{eq:deltadelta_full}\end{subequations}
\end{enumerate}
Equations (\ref{eq:diffusion_eq}) through (\ref{eq:flux_continuity_B_on_II}) represent a complete set determining $\delta \delta c^{(2)}_A(\mathbf{r} | {\mathbf r}_{I}, {\mathbf r}_{II})$ and $\delta \delta c^{(2)}_B(\mathbf{r} | {\mathbf r}_{I}, {\mathbf r}_{II})$.  Inhomogeneity in these equations is present only in boundary conditions (\ref{eq:flux_continuity_A_on_II}, \ref{eq:flux_continuity_B_on_II}), it comes from $\left. \delta c^{(1)}_A(\mathbf{r} | {\mathbf r}_{I}) \right|_{\mathbf{r} = \mathbf{r}_{II} + \mathbf{R}}$ and $\left. \delta c^{(1)}_B(\mathbf{r} | {\mathbf r}_{I}) \right|_{\mathbf{r} = \mathbf{r}_{II} + \mathbf{R}}$ -- the tails of concentration profiles created by particle $I$ in the region of particle $II$.  As long as particles are relatively far apart compared at their sizes, $\left| {\mathbf r}_{I} - {\mathbf r}_{II} \right| \gg R$, for the vicinity of particle $II$ we should write
\be \begin{split} \left.  \delta c^{(1)}_A(\mathbf{r} | {\mathbf r}_{I}) \right|_{\mathbf{r} = \mathbf{r}_{II} + \mathbf{R}} & \simeq  \delta c^{(1)}_A(\mathbf{r}_{II} | {\mathbf r}_{I}) + \\ & +  \mathbf{R} \cdot \nabla \delta c^{(1)}_A(\mathbf{r}_{II} | {\mathbf r}_{I}) \ , \end{split} \label{eq:linearized_tail} \ee
and similarly for $\left.  \delta c^{(1)}_B(\mathbf{r} | {\mathbf r}_{I}) \right|_{\mathbf{r} = \mathbf{r}_{II} + \mathbf{R}}$, $\left.  \delta c^{(1)}_A(\mathbf{r} | {\mathbf r}_{II}) \right|_{\mathbf{r} = \mathbf{r}_{I} + \mathbf{R}}$, and $\left.  \delta c^{(1)}_B(\mathbf{r} | {\mathbf r}_{II}) \right|_{\mathbf{r} = \mathbf{r}_{I} + \mathbf{R}}$.  We should plug this in the boundary conditions (\ref{eq:flux_continuity_A_on_II}, \ref{eq:flux_continuity_B_on_II}).  The inhomogeneous right hand sides of these boundary conditions then become sums of two terms each, one proportional to the value of concentration created by the seond particle in the center of the first one (and vice versa), $\delta c^{(1)}_A(\mathbf{r}_{I} | {\mathbf r}_{II})$  and $\delta c^{(1)}_B(\mathbf{r}_{I} | {\mathbf r}_{II})$, and another proportional to corresponding gradients, $\nabla \delta c^{(1)}_A(\mathbf{r}_{I} | {\mathbf r}_{II})$ etc.  Accordingly, we can split the solution for $\delta \delta c^{(2)}_A(\mathbf{r} | {\mathbf r}_{I}, {\mathbf r}_{II})$ in the vicinity of particle $I$ into a sum of two terms, one proportional to concentrations $\delta c^{(1)}_A(\mathbf{r}_{I} | {\mathbf r}_{II})$ and $\delta c^{(1)}_B(\mathbf{r}_{I} | {\mathbf r}_{II})$, and another to gradients, $\nabla \delta c^{(1)}_A(\mathbf{r}_{I} | {\mathbf r}_{II})$ and $\nabla \delta c^{(1)}_B(\mathbf{r}_{I} | {\mathbf r}_{II})$.  The former solution is spherically symmetric around particle $I$ and, therefore, contributes nothing to the force exerted on this particle.  The latter solution is proportional to gradient and, therefore, its contribution to the force decays with distance as $1/r^3$ and should be neglected as such.

Thus, we have justified the idea that every particle is driven, to the first approximation, by the concentration gradient generated by the other particle.  ``Interference'' effect does exist, as the concentration field created by one sphere is affected by the other sphere, but it becomes relevant only in the sub-leading term with respect to $r/R \ll 1$ and it is neglected in this work.

\end{document}